%% file: main.tex
\documentclass[sigconf]{acmart}

\usepackage{xspace}
\usepackage{longtable}
\usepackage{array}
\usepackage{graphicx}
\usepackage{multirow}
\usepackage{tabularx}
\usepackage{enumitem}

\AtBeginDocument{%
  }

\copyrightyear{2026}
\acmYear{2026}
\setcopyright{cc}
\setcctype{by}
\acmConference[Q-SE '26]{17h International Workshop on Quantum Software Engineering }{April 12--18, 2026}{Rio de Janeiro, Brazil}
\acmBooktitle{17h International Workshop on Quantum Software Engineering (Q-SE '26), April 12--18, 2026, Rio de Janeiro, Brazil}
\acmPrice{}
\acmDOI{10.1145/3786150.3788614}
\acmISBN{979-8-4007-2383-4/2026/04}

\begin{document}

\title{Industry Expectations and Skill Demands in Quantum Software Testing}


\author{Ronnie de Souza Santos}
\email{ronnie.desouzasantos@ucalgary.ca}
\orcid{0000-0003-3235-6530}
\affiliation{%
  \institution{University of Calgary}
  \city{Calgary}
  \state{Alberta}
  \country{Canada}}

\author{Maria Teresa Baldassarre}
\email{mariateresa.baldassarre@uniba.it}
\orcid{}
\affiliation{%
  \institution{University of Bari}
  \city{Bari}
  \country{Italy}
}

\author{César França}
\email{franssa@ufrpe.br}
\affiliation{%
  \institution{University of Bari}
  \city{Bari}
  \country{Italy}
}

\renewcommand{\shortauthors}{XXX et al.}
\begin{abstract}
\input{abstract}
\end{abstract}

\begin{CCSXML}
<ccs2012>
 <concept>
  <concept_id>00000000.0000000.0000000</concept_id>
  <concept_desc>Do Not Use This Code, Generate the Correct Terms for Your Paper</concept_desc>
  <concept_significance>500</concept_significance>
 </concept>
 <concept>
  <concept_id>00000000.00000000.00000000</concept_id>
  <concept_desc>Do Not Use This Code, Generate the Correct Terms for Your Paper</concept_desc>
  <concept_significance>300</concept_significance>
 </concept>
 <concept>
  <concept_id>00000000.00000000.00000000</concept_id>
  <concept_desc>Do Not Use This Code, Generate the Correct Terms for Your Paper</concept_desc>
  <concept_significance>100</concept_significance>
 </concept>
 <concept>
  <concept_id>00000000.00000000.00000000</concept_id>
  <concept_desc>Do Not Use This Code, Generate the Correct Terms for Your Paper</concept_desc>
  <concept_significance>100</concept_significance>
 </concept>
</ccs2012>
\end{CCSXML}

\ccsdesc[500]{Software and its engineering~Software creation and management~Software development process management}
\keywords{Quantum Software Engineering, Software Testing, Software Professionals}


\maketitle

\newcommand{\searchStrings}{3\xspace}
\newcommand{\topResults}{100\xspace}
\newcommand{\totalResults}{400\xspace}
\newcommand{\initialResults}{100\xspace}
\input{introduction}
\input{background}

\input{method}
\input{results}
\input{discussion}
\input{conclusion}

\bibliographystyle{ACM-Reference-Format}
\bibliography{bibliography}

\appendix

\end{document}

%% file: abstract.tex
Quantum software testing introduces new challenges that differ fundamentally from those in classical software engineering. {\textit{Aims:}} This study investigates how the quantum software industry defines testing roles and what skills are expected from professionals in these positions. {\textit{Method:}} We analyzed 110 job postings from organizations involved in quantum software and hardware development, identifying activities, competencies, and skill requirements related to testing. {\textit{Results:}} The findings show that testing in quantum contexts combines traditional software quality assurance with experimental validation, emphasizing calibration, control, and hybrid quantum–classical verification. Employers seek professionals who integrate programming and automation expertise with quantum-specific technical knowledge and interdisciplinary collaboration skills. {\textit{Conclusions:}} Quantum software testing remains at an early but rapidly evolving stage that bridges software engineering and experimental physics, highlighting the need for educational and research efforts that align testing practices with industrial realities.

%% file: introduction.tex
\section{Introduction}
\label{sec:introduction}

Software testing is a fundamental software engineering activity that plans, designs, executes, and evaluates tests to uncover defects and improve system reliability, usability, and performance \cite{whittaker2002software}. It is also a socio-technical practice, requiring analytical reasoning and tool proficiency alongside collaboration across roles \cite{kassab2017software, kochhar2019practitioners, Pardo2010}. Effective testers therefore combine hard skills (e.g., programming, automation, debugging, and domain knowledge) with soft skills (e.g., communication, teamwork, adaptability, and critical thinking), while ongoing trends such as automation, AI, and global collaboration continue to reshape testing work without eliminating the need for this balance \cite{florea2018software, florea2019skills, sanchez2020beyond, valle2023soft, florea2019global, gurcan2022evolution, lima2023software}.

Quantum computing reshapes software testing by altering how programs are specified, executed, and verified \cite{leite2025testing, garcia2023quantum, murillo2025quantum}. Quantum behavior is inherently probabilistic and only indirectly observable due to measurement-induced collapse, and it is further affected by noise and decoherence \cite{de2024quantum, paltenghi2024survey, SantaBarletta2025, miranskyy2021testing, ali2023quantum, garcia2023quantum}. These constraints challenge classical testing assumptions and motivate quantum-specific techniques (e.g., for unitary transformations) and adaptations of test generation, mutation testing, metamorphic relations, and coverage criteria to quantum circuits and hybrid quantum--classical systems \cite{ali2024quantum, leite2025testing, ali2024quantum, garcia2023quantum, paltenghi2024survey, zappin2025challenges}.

Although research on quantum software testing is expanding, there is still limited evidence on how these advances translate into industrial practice \cite{paltenghi2024survey}. As organizations rapidly build quantum teams, their expectations for testing-related roles—spanning prior testing experience, tool familiarity, and interpersonal competencies—remain unclear \cite{zappin2025challenges, de2024quantum, leite2025testing}. To help align education, training, and professional development in Quantum Software Engineering, this study investigates industry demand and characterizes professional expectations for quantum software testing roles.

To achieve this, we collected and analyzed 110 publicly available job postings\footnote{Data available at \url{https://figshare.com/s/6ac4583b0ec96e545c4d}} for positions that explicitly focus on testing or include testing responsibilities in the context of quantum software development. Based on this, we addressed the following research questions: \textbf{RQ1.} What testing-related activities and competencies are requested in job postings for quantum software engineers?
\textbf{RQ2.} What hard and soft skills are expected from professionals responsible for testing quantum systems? Our contributions are threefold: (1) we provide an empirical overview of the testing expectations expressed in the quantum software job market, (2) we map the relationship between industry expectations and current research on quantum software testing, and (3) we discuss implications for education and professional training that can guide curriculum design and workforce preparation.

The remainder of this paper is organized as follows. Section~\ref{sec:background} reviews software testing skills and quantum software testing. Section~\ref{sec:method} describes the data collection and qualitative analysis. Section~\ref{sec:findings} presents the testing-related demands and skills identified in the job postings, and Section~\ref{sec:discussion} discusses implications for research, education, and practice. Section~\ref{sec:conclusions} concludes and outlines future work.

%% file: background.tex
\section{Background}
\label{sec:background}
This section introduces the key foundations of the study by discussing the professional competencies required for software testing and the characteristics and challenges that define quantum software testing.

\subsection{Hard and Soft Skills for Software Testing}

Early work already framed testing as a rigorous yet creative activity requiring structured thinking \cite{whittaker2002software}. As software systems and development organizations have grown more complex and distributed, testers are increasingly expected to combine technical proficiency with human-centered competencies to sustain effective quality assurance in dynamic environments \cite{florea2019skills, valle2023soft, kassab2017software, kochhar2019practitioners}.

Empirical studies show that effective testing combines complementary hard and soft skills \cite{valle2023soft}. Hard skills include the technical knowledge to implement, automate, and evaluate tests (e.g., programming, test frameworks, and CI tooling) \cite{florea2019global, florea2019skills}. Soft skills cover interpersonal and cognitive abilities—such as communication, teamwork, and critical thinking—that help testers navigate ambiguity and coordination in large projects \cite{sanchez2020beyond, florea2018software, valle2023soft}. Practitioner surveys suggest that technical expertise alone is insufficient; collaboration and priority negotiation are therefore central to the tester role \cite{florea2018software, sanchez2020beyond}.

Educational programs and professional training increasingly need to address the dual technical and human dimensions of testing, as curricula and job-market studies report persistent mismatches between academic preparation and industry expectations \cite{lima2023software, valle2023soft, florea2019skills, sanchez2020beyond, gurcan2022evolution}. Employers therefore prioritize hybrid profiles that combine automation competence with the ability to interpret results, reason about system behavior, and communicate effectively in cross-functional settings \cite{lima2023software, valle2023soft, gurcan2022evolution, florea2019global}. Overall, the literature suggests that advances in testing depend not only on technological innovation but also on cultivating such blended skill sets that integrate computational precision with human judgment \cite{florea2018software, florea2019global, gurcan2022evolution, valle2023soft, sanchez2020beyond, kochhar2019practitioners}.

\subsection{Quantum Software Testing}

Quantum computing introduces testing challenges that differ from classical settings because program behavior is probabilistic and internal states are only indirectly observable, which weakens traditional assumptions about correctness, determinism, and reproducibility \cite{rieffel2000introduction, miranskyy2021testing, de2024quantum, garcia2023quantum, ali2023quantum}. As a result, debugging and verification often require new analytical techniques and tool support tailored to quantum execution and measurement \cite{paltenghi2024survey, rieffel2000introduction}.

The literature on quantum software testing has expanded rapidly, proposing methods that adapt classical testing principles to probabilistic quantum behavior, including oracle- and assertion-based verification, metamorphic relations, and mutation analysis \cite{paltenghi2024survey, Barletta2024, ali2024quantum, garcia2023quantum}. Surveys and mappings organize these techniques by the subproblems they target (e.g., correctness, performance, and hardware-related analysis) and commonly highlight the need for statistical reasoning and hybrid quantum--classical execution \cite{paltenghi2024survey, leite2025testing}. Despite this conceptual progress, empirical evidence remains limited on how such methods are implemented in practice and aligned with industrial workflows \cite{ali2023quantum, ali2024quantum, garcia2023quantum, paltenghi2024survey, leite2025testing, zappin2025challenges}.

Recent evidence suggests that quantum software testing remains formative: empirical studies report that testing is largely manual and often relies on simulation or small-scale circuit checks, supported by ad-hoc debugging and visualization \cite{zappin2025challenges, Serrano2024}. Specialized testing frameworks are still underutilized due to limited awareness and integration challenges, reinforcing a gap between academic tooling and industrial adoption \cite{zappin2025challenges, garcia2023quantum}. Practitioners therefore highlight the need for scalable, usable, and well-documented solutions, alongside clearer professional standardization and training pathways to enable broader industrial uptake \cite{de2024quantum, leite2025testing}.

%% file: method.tex
\section{Method}
\label{sec:method}

To investigate the testing-related expectations and required skills for professionals in quantum software engineering, we conducted a systematic analysis of job postings. Job advertisements are widely used in empirical software engineering research because they offer near real-time evidence of workforce demand and detailed descriptions of professional roles \cite{papoutsoglou2019extracting, lecca2025curious, Pardo201159}. This approach allowed us to identify how testing activities, competencies, and qualifications are represented within the emerging quantum software industry.

\subsection{Data Collection}

The data collection took place between August and October 2025, and followed three complementary strategies designed to ensure coverage and reliability:

\begin{enumerate}
\item \textbf{Job Portals:} We performed keyword-based searches on major job portals (LinkedIn, Indeed,  Glassdoor, Zip Recruiter, Jooble, Flexjobs, JobLeads, Quantum Jobs US and Workday), retrieving a total of \textbf{9,818 postings}. These results included a wide range of positions across the quantum technology ecosystem. We used "Quantum" as the main keyword, and complemented the search with specific technologies keywords (Qiskit, Cirq, Q\#, QDK, PennyLane, QuTiP, D-Wave, OceanSDK, Quill, PyQuill, OpenQuasm).

\item \textbf{Company Websites:} In parallel, we examined the official career pages of companies known for quantum software or hardware development (e.g., IBM, Google, Microsoft, Rigetti, and Xanadu). This strategy added \textbf{352 postings} not indexed by job portals.

\item \textbf{Data Validation:} We removed duplicates during collection and conducted peer validation to confirm the consistency of the retrieved postings and ensure that all belonged to the software domain.
\end{enumerate}

\subsection{Filtering and Selection}

From the 10,170 retrieved postings, we applied successive filters to identify those relevant to the focus of this study:

\begin{itemize}
\item \textbf{Filter 1 – Domain Relevance:} We first selected postings explicitly associated with software engineering or development within the context of quantum computing. This filter reduced the dataset to \textbf{128 postings} (about 1.2\% of the total).
\item \textbf{Filter 2 – Testing Focus:} We then retained only postings that either (a) described software testing positions or (b) explicitly mentioned testing-related activities such as verification, validation, debugging, or quality assurance. This yielded \textbf{110 postings}.
\item \textbf{Filter 3 – Completeness:} Finally, we checked for postings lacking sufficient descriptive content (e.g., empty descriptions or missing skill requirements). No posts were excluded in this phase.
\end{itemize}

The resulting dataset contained job postings from multiple geographic regions and organizational types, including start-ups, research laboratories, and large technology firms. This diversity ensured a representative view of industrial expectations for quantum software professionals.

\subsection{Data Analysis and Synthesis}
The analysis combined descriptive and thematic approaches to provide both quantitative and qualitative insights. Descriptive statistics summarized the frequency of testing-related mentions, required skills, and professional roles, offering a general overview of the patterns observed in the job market \cite{george2018descriptive}. Thematic synthesis was then conducted to interpret the nature of testing expectations in the context of quantum software. Following guidelines for qualitative synthesis in software engineering \cite{cruzes2011recommended}, we coded textual segments describing testing activities, desired competencies, and professional attributes. The codes were iteratively refined into higher-level categories that reflected testing tasks, challenges, and skill types identified in prior literature on quantum software testing (e.g., test input generation, circuit simulation, probabilistic verification, metamorphic testing, and hybrid quantum–classical validation). This process enabled the integration of quantitative trends with conceptual understanding, highlighting how industrial job descriptions operationalize the testing concepts, challenges, and skill demands emphasized in the research literature on quantum software testing.

%% file: results.tex
\section{Results}
\label{sec:findings}
This section summarizes the findings from 110 job postings related to quantum software development. We first outline the general demographics of the postings, then examine those explicitly focused on testing, and finally explore other roles that require testing or validation skills as part of their duties. The job postings analyzed in this study are available at \url{https://figshare.com/s/6ac4583b0ec96e545c4d}. Across this section, we refer to the postings using their coding in the available material. 

\subsection{Demographics}
From the 110 job postings identified, five were exclusively focused on quantum software testing, meaning that these positions explicitly sought professionals whose primary role was to design, execute, and maintain tests for quantum systems. The remaining 105 postings corresponded to other positions in quantum software development that required candidates to demonstrate testing expertise as part of their responsibilities. Among these postings, most referred to software engineering and research roles that incorporated testing within broader development activities. Common titles included Software Engineer, Quantum Software Engineer, Research Software Engineer, and Control and Calibration Engineer.

These positions emphasized testing-related responsibilities, including verifying algorithmic correctness, validating hardware–software integration, and ensuring the reliability of hybrid quantum–classical systems. A smaller group of senior and leadership roles, including Engineering Productivity Manager and Site Reliability Engineer, highlighted testing as part of quality assurance and workflow automation. These postings indicate that testing skills are viewed as essential across many areas of quantum software development.

Another aspect of the positions is related to their global distribution. We observed that the job postings were concentrated in regions with established quantum technology activity, primarily in North America and Europe. The United States accounted for 83 postings, followed by the United Kingdom with 6, France with 5, and Australia with 4. Additional postings were identified in India (3), Canada (2), and Italy (2), with single entries from Austria, Denmark, Germany, Switzerland, and Ireland. This distribution reflects the concentration of both industrial and academic quantum computing initiatives in these regions.

\subsection{Quantum Software Testing Key Activities in Industry}

To understand how quantum software testing is practiced in industrial settings, we compared the activities described in the literature \cite{ali2023quantum, ali2024quantum, paltenghi2024survey, leite2025testing, zappin2025challenges, miranskyy2021testing,Patón-Romero2019180} with the responsibilities outlined in the job postings explicitly dedicated to testing roles. The academic sources identify a range of activities that combine traditional quality assurance methods with quantum-specific validation techniques. These include automation of calibration and measurement routines, hardware-in-the-loop and device characterization, quantum error-correction and fidelity evaluation, hybrid quantum–classical interface testing, and noise-aware or probabilistic verification. Table \ref{tab:QuantumTestingActivities} summarize these findings.

Our analysis of the five industry postings reveals that three of them (AD-Test-001, AD-Test-002, and AD-Test-003) required professionals to perform quantum-specific testing activities, including hardware-in-the-loop validation, device characterization, and automated calibration and measurement routines. Among these, AD-Test-002 explicitly mentioned familiarity with quantum control and pulse-level programming, while AD-Test-003 focused on testing quantum error-correction systems and assessing fidelity across different hardware backends. In addition, hybrid quantum–classical validation appeared in AD-Test-003 and AD-Test-004, and data-oriented validation was emphasized in AD-Test-001, which required organizing and managing experimental test and measurement data.

Other activities discussed in the literature were not yet reflected in any of the analyzed postings. These include noise-aware statistical testing, probabilistic oracles, metamorphic and mutation-based testing, and entanglement or teleportation verification. The absence of these advanced testing techniques suggests that while industrial practice has already integrated experimental and system-level testing for quantum devices, it has not yet adopted some of the validation strategies that dominate academic research.

\input{tables/testing-focused}

\subsection{Skills Required for Quantum Software Testing Roles}
\label{sec:skills}

After identifying the key testing activities, we examined the professional skills required in the job postings dedicated to quantum software testing. This analysis focused on the technical and interpersonal competencies that organizations expect from professionals performing testing and validation in quantum contexts. The postings consistently differentiated between technical (hard) and interpersonal (soft) abilities, showing that quantum testing demands expertise that bridges software engineering, experimental physics, and data-driven analysis. \\

\noindent \textbf{Hard Skills.}
Across the five postings, employers emphasized a combination of general software testing skills and quantum-specific technical abilities. Common requirements included proficiency in programming and scripting languages such as Python, MATLAB, and C++, experience with automated testing frameworks (e.g., Pytest, Selenium, Postman), and familiarity with continuous integration environments using Git, Jenkins, and Docker. These represent the traditional software engineering competencies expected in advanced QA and testing roles. However, several skills were identified as exclusive to quantum software testing. These were explicitly related to the control, calibration, and validation of quantum hardware and hybrid systems. AD-Test-001 and AD-Test-002 required professionals to develop and automate calibration and measurement routines, integrating control and readout tools for qubit operation. AD-Test-002 further demanded familiarity with pulse-level programming and quantum control software, while AD-Test-003 focused on testing quantum error-correction mechanisms and fidelity evaluation across multiple hardware backends. In addition, AD-Test-003 and AD-Test-004 required verification of hybrid quantum–classical interfaces, and AD-Test-001 emphasized data-oriented validation by organizing and managing experimental test data. These findings indicate that, beyond traditional programming and automation, quantum software testing roles increasingly require knowledge of experimental systems, calibration automation, and hybrid control architectures—competencies that are distinct from those observed in conventional software testing. \\

\input{tables/hard-skills-testing}

\noindent \textbf{Soft Skills.} The soft skills described in the postings reflected many of the same categories found in traditional software testing, including collaboration, communication, analytical reasoning, and adaptability. However, the context in which these skills are applied in quantum software testing is substantially different, given the interdisciplinary and experimental nature of the work. Interdisciplinary collaboration was a recurring theme across all postings and extended beyond coordination among developers and QA engineers to include physicists, hardware engineers, and control specialists. Testing professionals were expected to interact with laboratory teams and participate in activities that bridge software validation and experimental calibration, as observed in AD-Test-001, AD-Test-002, and AD-Test-004. This collaborative dynamic highlights that testing in quantum systems occurs at the interface of software and physical experimentation.

Communication across disciplines was also emphasized, particularly in AD-Test-002 and AD-Test-003, where professionals were expected to report and document results in ways that could be understood by both software engineers and physicists. Unlike traditional QA communication, which focuses primarily on technical reporting, these roles required translating between mathematical models, physical observations, and system behavior to ensure mutual understanding across diverse teams. Analytical reasoning under uncertainty was another skill emphasized across the postings. Debugging in quantum systems differs from classical contexts because outcomes are probabilistic and sometimes only partially observable. As reflected in AD-Test-002, AD-Test-003, and AD-Test-004, analytical reasoning is therefore less about identifying deterministic defects and more about interpreting noisy experimental results and diagnosing system variability.

Finally, adaptability and scientific curiosity appeared as essential attributes in positions such as AD-Test-002 and AD-Test-003. In classical QA, adaptability typically refers to learning new frameworks or tools. In quantum testing, it extends to understanding evolving physical principles and rapidly advancing technologies, such as emerging qubit architectures and quantum error-correction methods. Overall, the soft skills described across these postings reveal that while collaboration, communication, analytical reasoning, and adaptability remain essential, their application in quantum contexts is defined by scientific integration, probabilistic reasoning, and continuous learning in an environment where experimentation and computation coexist.

\input{tables/soft-skills-testing}

\subsection{Testing-Related Tasks in Broader Quantum Engineering Roles}

Beyond the postings explicitly dedicated to testing, our dataset includes numerous positions with broader titles such as \textit{Software Engineer}, \textit{Quantum Software Engineer}, \textit{Research Software Engineer}, and \textit{Control and Calibration Engineer}. Although these roles are not formally labeled as testing positions, their descriptions reveal that professionals in these categories are nonetheless responsible for performing several testing-related tasks as part of their daily activities. These activities extend traditional responsibilities in software development and system engineering to encompass the validation of hybrid quantum–classical architectures, calibration automation, hardware–software integration, and performance benchmarking.

In our analysis of the 105 \textit{AD-Test-Support} postings, we observed a consistent demand for testing and validation capabilities across engineering and research-oriented positions. The most frequent and mature activities—those appearing in a high or medium number of postings—include:

\begin{itemize}

\item \textbf{Hardware-in-the-loop and device validation (19 postings)} — This activity appears most frequently across control, embedded, and hardware–software integration roles. Professionals are expected to develop and execute tests that connect control software with physical devices, verify the behavior of signal generation and measurement systems, and validate firmware and hardware performance. Typical responsibilities include integrating instruments into testbeds, implementing automated measurement workflows, and conducting device characterization through closed-loop testing.

\item \textbf{Quantum testbed operation and system-level experimentation (25 postings)} — System-level validation represents a central aspect of industrial practice. Engineers are expected to operate and maintain experimental testbeds, coordinate software–hardware integration, and execute full-stack experiments that assess end-to-end system performance. This activity typically involves debugging distributed control systems, verifying synchronization across subsystems, and analyzing test outcomes under real laboratory or deployment conditions.

\item \textbf{Automation of calibration and measurement routines (18 postings)} — Frequently mentioned in both software and hardware development roles, this task involves designing scripts and frameworks that automate device calibration, signal optimization, and measurement collection. Engineers are asked to build autonomous routines for calibration and alignment, ensuring system stability and repeatability across multiple experimental cycles.

\item \textbf{Integration of test data and experimental logging (19 postings)} — A growing number of positions emphasize data-driven validation practices. Candidates are required to design and maintain data pipelines for test and measurement results, implement logging and observability frameworks, and ensure traceability of experiments through systematic data management, analysis, and reporting.

\item \textbf{Quantum control and pulse-level testing (19 postings)} — These tasks appear primarily in positions focused on hardware control and firmware development. Professionals are required to test and validate pulse sequences, control waveforms, and timing configurations, often using low-level programming interfaces or domain-specific control languages. The emphasis is on ensuring synchronization, timing accuracy, and robustness of control operations across experimental environments.

\item \textbf{Hybrid quantum–classical interface validation (18 postings)} — As hybrid computation becomes the prevailing industrial model, engineers are frequently responsible for testing the interfaces between classical software stacks and quantum backends. This includes validating data flow between APIs, ensuring interoperability across heterogeneous environments, and assessing the reliability of hybrid execution pipelines used in cloud-based quantum services.

\item \textbf{Quantum error-correction and fidelity testing (12 postings)} — Although less common, several postings describe responsibilities related to the testing and verification of quantum error-correction protocols and performance benchmarking. Tasks include developing test cases for fidelity evaluation, implementing diagnostic tools for system health monitoring, and integrating fault-tolerant control procedures within larger testing frameworks.

\end{itemize}

These findings indicate that testing-related tasks are deeply embedded in engineering and research positions across the quantum computing ecosystem, even when these roles are not explicitly designated as testing positions. Industrial practice thus emphasizes \textit{hardware-coupled, automation-intensive, and integration-oriented} testing, reflecting a pragmatic and system-level approach to ensuring reproducibility, stability, and reliability in quantum technologies.

Regarding testing skills, in the non-dedicated testing positions, employers emphasized a blend of software engineering fundamentals and specialized technical competencies that enable testing, validation, and quality assurance in quantum development environments. Traditional software testing proficiencies remained essential, including programming skills in Python, C++, and Rust, the use of automated testing frameworks and scripting environments, and experience with continuous integration and deployment tools such as Git, Jenkins, Docker, and Kubernetes. These represent the core verification practices that underpin large-scale software reliability across hybrid infrastructures.

At the same time, several postings introduced hard skills that extend these traditional competencies into quantum-specific contexts. Candidates were expected to apply software validation principles to experimental and embedded systems through hardware-in-the-loop testing, firmware debugging, and FPGA integration. Many roles required experience in automated calibration pipelines and measurement scripting, particularly for optimizing control parameters and ensuring device stability. Others highlighted expertise in pulse-level programming, signal processing, and waveform verification as part of quantum control validation. Continuous integration and monitoring were also reframed for experimental settings, with employers seeking familiarity with observability tools such as Prometheus and Grafana for logging and diagnosing test outcomes. Finally, several postings requested knowledge of hybrid API testing and system-level verification across quantum–classical boundaries.

These patterns reveal that even in positions not labeled as testing roles, professionals are required to demonstrate strong capabilities in automation, integration testing, and experimental validation. In quantum engineering practice, testing hard skills therefore converges around hybrid software–hardware proficiency, experimental calibration and control, and the adaptation of continuous integration and debugging techniques to laboratory and quantum computing environments.

%% file: tables/testing-focused.tex
\begin{table}[ht]
\caption{Quantum-Specific Testing Techniques Identified in Industry Job Postings}
\label{tab:QuantumTestingActivities}
\scriptsize
\centering
\begin{tabular}{p{1.7cm} p{1.3cm} p{4.7cm}}
\toprule
\textbf{Quantum Testing Technique} & \textbf{Job Ad (Code)} & \textbf{Evidence (Explicit Excerpts from Job Postings)} \\
\midrule

Hardware-in-the-loop validation and device characterization 
& AD-Test-001, AD-Test-002, AD-Test-003 
& “Develop software to control a variety of signal generation and measurement hardware.” (AD-Test-001) \newline
“Rigorously test our innovative software and hardware solutions and measure performance.” (AD-Test-003) \\
\midrule

Automated calibration and measurement routines 
& AD-Test-001, AD-Test-002 
& “Develop software for the autonomous execution of complex sequences of experiments, measurements, and calibrations.” (AD-Test-001) \newline
“Integrate control, readout, and calibration tools into our product stack.” (AD-Test-002) \\
\midrule

Quantum control and pulse-level testing 
& AD-Test-002 
& “Familiarity with pulse-level programming languages or quantum control software.” (AD-Test-002) \\
\midrule

Quantum error-correction and fidelity testing 
& AD-Test-003 
& “Testing of Riverlane’s quantum error correction (QEC) systems.” \newline
“Measure performance and identify integration issues early on.” (AD-Test-003) \\
\midrule

Hybrid quantum–classical interface validation 
& AD-Test-003, AD-Test-004 
& “Testing software and hardware solutions across different environments and for multiple types of quantum computers.” (AD-Test-003) \newline
“Collaborate with software, hardware, and quantum physics teams to ensure high test coverage and product quality.” (AD-Test-004) \\

\bottomrule
\end{tabular}
\vspace{-0.5cm}
\end{table}

%% file: tables/hard-skills-testing.tex
\begin{table}[ht]
\caption{Hard Skills in Quantum Software Testing Positions}
\label{tab:QuantumExclusiveHardSkills}
\scriptsize
\centering
\begin{tabular}{p{1.7cm} p{1.3cm} p{4.7cm}}
\toprule
\textbf{Quantum-Exclusive Skill} & \textbf{Job Ad (Code)} & \textbf{Evidence (Explicit Excerpts from Job Postings)} \\
\midrule

Quantum control and pulse-level programming 
& AD-Test-002 
& “Familiarity with pulse-level programming languages or quantum control software.” (AD-Test-002) \\
\midrule

Quantum hardware operation and characterization 
& AD-Test-001, AD-Test-002 
& “Develop software to control a variety of signal generation and measurement hardware.” (AD-Test-001) \newline
“Assemble, operate, and maintain a quantum computing testbed platform.” (AD-Test-002) \\
\midrule

Quantum error-correction and fidelity evaluation 
& AD-Test-003 
& “Testing of Riverlane’s quantum error correction (QEC) systems.” \newline
“Measure performance and identify integration issues early on.” (AD-Test-003) \\
\midrule

Hybrid quantum–classical system integration 
& AD-Test-003, AD-Test-004 
& “Testing software and hardware solutions across different environments and for multiple types of quantum computers.” (AD-Test-003) \newline
“Collaborate with software, hardware, and quantum physics teams to ensure high test coverage and product quality.” (AD-Test-004) \\
\midrule

Automation of calibration and measurement routines 
& AD-Test-001, AD-Test-002 
& “Develop software for the autonomous execution of complex sequences of experiments, measurements, and calibrations.” (AD-Test-001) \newline
“Integrate control, readout, and calibration tools into our product stack.” (AD-Test-002) \\

\bottomrule
\end{tabular}
\end{table}

%% file: tables/soft-skills-testing.tex
\begin{table}[ht]
\caption{Soft Skills in Quantum Software Testing Positions}
\label{tab:SoftSkillsQuantumTesting}
\scriptsize
\centering
\begin{tabular}{p{1.7cm} p{1.3cm} p{4.7cm}}
\toprule
\textbf{Soft Skill} & \textbf{Job Ad (Code)} & \textbf{Evidence (Explicit Excerpts from Job Postings)} \\
\midrule

Interdisciplinary collaboration 
& AD-Test-001, AD-Test-002, AD-Test-004 
& “Work closely with experimental physicists to ensure software enables laboratory experimentation.” (AD-Test-001) \newline
“Collaborate with software and hardware teams to integrate control, readout, and calibration tools.” (AD-Test-002) \newline
“Collaborate with software, hardware, and quantum physics teams to ensure high test coverage and product quality.” (AD-Test-004) \\
\midrule

Communication across disciplines 
& AD-Test-002, AD-Test-003 
& “Excellent collaboration, communication, and documentation skills.” (AD-Test-002) \newline
“Excellent communication skills, both written and verbal.” (AD-Test-003) \\
\midrule

Analytical reasoning under uncertainty 
& AD-Test-002, AD-Test-003, AD-Test-004 
& “Strong troubleshooting and debugging skills.” (AD-Test-002) \newline
“Identify the underlying causes of test failures.” (AD-Test-003) \newline
“Strong analytical, debugging, and problem-solving skills.” (AD-Test-004) \\
\midrule

Adaptability and scientific curiosity 
& AD-Test-002, AD-Test-003 
& “Highly motivated, adaptable, and passionate about driving technical progress.” (AD-Test-002) \newline
“Curiosity to learn about new technologies, including quantum computing and quantum error correction.” (AD-Test-003) \\

\bottomrule
\end{tabular}
\end{table}

%% file: discussion.tex
\section{Discussion}
\label{sec:discussion}

In this section, we answer our research questions, relate the findings to the existing literature, and discuss their implications and limitations.

\subsection{Answering the Research Questions}

Our first research question asked {\textit{What testing-related activities and competencies are requested in job postings for quantum software engineers?}} 
The analysis shows that industrial practice combines traditional software testing with emerging quantum-specific validation. Classical testing remains present through automation, verification, and integration testing, but in quantum contexts these practices are extended to include hardware-in-the-loop validation, calibration automation, fidelity testing, and hybrid quantum–classical verification. The postings demonstrate that quantum testing is inseparable from experimentation, as professionals are expected to design and execute validation routines that connect software directly to physical devices. The prominence of activities such as device characterization, signal measurement, and control pulse verification indicates that testing operates at the intersection of software engineering and experimental physics. 

Several advanced testing strategies described in the academic literature, such as metamorphic testing, mutation analysis, probabilistic oracles, and noise-aware validation, were not present in the job postings. This difference suggests that industrial quantum testing remains largely system- and hardware-centered rather than oriented toward statistical or formal validation frameworks. Industrial practice therefore reflects an early but pragmatic stage of maturity, where testing focuses on reproducibility, calibration stability, and interface reliability rather than formal verification of quantum algorithms. These findings highlight that while academic research has proposed a wide array of testing techniques, industry adoption is currently concentrated on experimental control and integration testing.

Our second research question asked {\textit{What hard and soft skills are expected from professionals responsible for testing quantum systems?}} 
Employers expect testing professionals to demonstrate both general software quality assurance expertise and quantum-exclusive technical abilities. Hard skills include programming and scripting in Python, C++, and MATLAB, the use of automation frameworks and continuous integration pipelines, and experience with testing and data management tools. The distinctive competencies relate to quantum control and calibration, including pulse-level programming, device operation, error-correction testing, and hybrid interface verification. These requirements show that quantum testing roles demand cross-domain expertise that bridges coding, experimental design, and hardware control.

Soft skills mirror those emphasized in traditional software testing, including collaboration, communication, analytical reasoning, and adaptability, but their application occurs in more complex and interdisciplinary settings. Testing professionals must coordinate with physicists, engineers, and data scientists, translating experimental results into computational insights. Communication extends beyond reporting technical defects to mediating understanding between teams grounded in different scientific paradigms. Analytical reasoning also changes in character, as professionals must interpret noisy and probabilistic outcomes instead of deterministic failures. Adaptability becomes essential because quantum systems evolve rapidly and testing frameworks require continuous reconfiguration. Overall, these roles illustrate that quantum software testing is a deeply socio-technical endeavor that requires technical precision, cross-disciplinary fluency, and tolerance for uncertainty.

\subsection{Comparing Results with the Literature}
\label{sec:discussioncomparing}

Our findings align with prior research describing the interdisciplinary and probabilistic nature of quantum software testing. Similar to earlier observations that classical testing paradigms cannot be directly transferred to quantum systems \cite{miranskyy2021testing, ali2023quantum, garcia2023quantum}, our results show that industrial practice focuses on hybrid verification strategies and calibration-driven experimentation. The prevalence of testing activities involving hardware-in-the-loop, control pulses, and fidelity evaluation confirms that organizations are prioritizing the reliability of physical systems over purely algorithmic correctness, consistent with the system-level emphasis reported by \cite{leite2025testing} and \cite{zappin2025challenges}. These results reinforce that quantum software testing in industry remains closely tied to the operational needs of laboratory environments and device stability.

Our results also diverge from much of the current academic discourse. Whereas research on quantum testing emphasizes formalism through mutation operators, metamorphic relations, probabilistic oracles, and fault injection, industry postings rarely mention these activities. This divergence suggests a gap between the conceptual progress of research and its practical adoption. The reasons may include limited tool maturity, integration challenges, or the absence of standardized frameworks that connect academic prototypes and industrial workflows. Similar mismatches have been noted in other emerging domains of software engineering where theory advances faster than tool deployment \cite{garcia2023quantum, ali2024quantum}. In this sense, our study empirically confirms that quantum software testing remains at a formative stage marked by early experimentation rather than consolidated best practices.

Our findings also extend the literature on software testing skills. Previous studies have consistently identified the dual importance of hard and soft skills for effective testing \cite{florea2018software, sanchez2020beyond, valle2023soft}, and our results indicate that this duality persists in quantum contexts but with new interpretations. Technical skills now encompass quantum-specific programming and control, while soft skills expand to scientific collaboration and cross-domain communication. This aligns with recent observations that the success of quantum software projects depends on professionals who can mediate between software abstractions and physical experimentation \cite{zappin2025challenges, leite2025testing}. By empirically mapping these competencies, our results contribute to the ongoing discussion about what professional preparation is required for the emerging field of quantum software engineering.

\subsection{Implications}

From a research perspective, our findings suggest that quantum software testing should be studied as a hybrid discipline integrating software engineering, physics, and other domains. The gap between academic testing frameworks and industrial practice calls for more empirical work examining how testing is performed in real laboratory settings and how academic methods can be adapted for these environments. Longitudinal studies could trace the evolution of testing maturity as quantum technologies transition from prototypes to production systems. Future research may also focus on human and organizational factors, such as collaboration, training, and communication across disciplines, that shape testing success as much as algorithmic innovation.

From an educational and professional development perspective, our results indicate that training programs for quantum software engineers must go beyond introducing quantum algorithms and programming languages. Curricula should integrate testing and validation concepts tailored to quantum systems, emphasizing calibration automation, control integration, and hybrid debugging. Equally important, soft skills such as interdisciplinary communication, adaptability, and problem solving under uncertainty should be explicitly developed. Collaborations between universities and industry could ensure that students experience real testing environments and learn to navigate the challenges of experimental reproducibility and data-driven validation. These findings therefore point to a need for professional formation that mirrors the hybrid, socio-technical character of quantum software testing.

For practitioners and organizations, our results highlight that testing responsibilities are widely distributed across engineering roles. Even when positions are not labeled as testing, professionals are expected to perform validation tasks that ensure system reliability. Organizations may therefore consider formalizing testing as a shared responsibility within quantum development pipelines. Establishing clearer testing standards, shared toolchains, and knowledge exchange mechanisms between researchers and engineers could accelerate the maturation of testing practices and improve reproducibility across the ecosystem.

\subsection{Threats to Validity}
The \textbf{credibility} of the findings is influenced by reliance on job postings as data sources, since advertisements may provide incomplete, simplified, or aspirational descriptions and require researcher interpretation. This limitation was partially mitigated by sampling postings from multiple organizations and cross validating reported responsibilities with existing literature. However, some postings include activities more closely related to quantum hardware engineering, which may bias the identified testing skill set toward hardware optimization rather than software oriented verification and validation. The \textbf{transferability} of the results is constrained by the geographic concentration of the sample in North America and Europe and by the rapidly evolving nature of job advertisements, which capture a time specific snapshot of an emerging industry. \textbf{Reflexivity} was explicitly considered, as the authors’ backgrounds in software engineering and research may have influenced how testing related responsibilities were interpreted. Finally, \textbf{rigor} was supported through systematic coding, comparison across postings, and triangulation with current literature, providing a transparent and replicable analysis based on publicly available data.

%% file: conclusion.tex
\section{Conclusions and Future Work}
\label{sec:conclusions}

This study analyzed 110 job postings to characterize how testing is defined and what skills are required in quantum software engineering. The results indicate that testing in this domain combines classical quality assurance practices with experimental validation, with emphasis on calibration, control, and hybrid quantum–classical verification. Employers seek professionals who combine programming and automation capabilities with quantum-specific knowledge, alongside collaboration, communication, and analytical reasoning across disciplinary boundaries. The findings indicate that required skills are not static and are likely to change as quantum technologies evolve. Current roles emphasize close interaction with hardware, experimental validation, and low-level system understanding, reflecting the early stage of industrial adoption. As quantum platforms mature, testing practices are likely to shift toward higher levels of abstraction, greater automation, and more standardized verification workflows. This evolution has direct implications for curriculum design and certification frameworks, which should be designed to accommodate changing skill profiles over time. Educational programs may need to balance enduring software engineering fundamentals with emerging quantum-specific competencies, while certification frameworks should remain adaptable to technological change rather than codifying a fixed set of skills. This study opens several avenues for follow-up research. First, a longitudinal replication (e.g., re-collecting postings in 12–24 months using the same protocol) would allow us to quantify how required testing skills evolve as platforms mature, highlighting emerging and declining competencies and the pace at which testing moves up the abstraction stack. Second, given that the dataset already enables richer exploration, future work can conduct deeper comparative analyses across regions, company types, and role seniority to identify statistically supported differences. Third, a more formal comparison with classical software testing roles—using established competency models from software testing/QA and mapping overlaps and deltas—would clarify which aspects of quantum testing reflect continuity with traditional practice and which are genuinely distinctive. Finally, future studies should explicitly classify skills by quantum software stack and framework, distinguishing generalizable testing competencies from those that are technology- or vendor-specific; this would better inform curriculum design and certification frameworks by separating enduring fundamentals from rapidly shifting, platform-bound skills.

\section{Acknowledgments}
\label{sec:acknowledgments}

This work has been partially supported by the following projects: "QUASAR: QUAntum software engineering for Secure, Affordable, and Reliable systems", grant 2022T2E39C, under the PRIN 2022 MUR program funded by the EU - NGEU; SERICS (PE00000014) under the MUR National Recovery and Resilience Plan funded by the European Union – NextGenerationEU; Patto territoriale "Sistema universitario pugliese" – CUP F61B23000370006